\documentstyle[aaspptwo]{article}
\newcommand{\dif}[2]{{\partial #1 \over \partial #2}}
\newcommand{\fracd}[2]{{\displaystyle{#1} \over \displaystyle{#2}}}

\eqsecnum
\received{November 7 1994}
\slugcomment{submitted to the Astrophys. J.}

\begin{document}

\title{Collapse and Fragmentation of Cylindrical Magnetized Clouds.~II.\\
Simulation with Nested Grid Scheme}

\author{Kohji Tomisaka\altaffilmark{1}}
\affil{Faculty of Education, Niigata University,
    8050 Ikarashi-2, Niigata 950-21, Japan}
\altaffiltext{1}{E-mail: tomisaka@ed.niigata-u.ac.jp.}

\begin{abstract}
Fragmentation process in a cylindrical magnetized cloud is studied with
 the nested grid method.
The nested grid scheme use 15 levels of grids with different spatial
 resolution overlaid subsequently, which enables us to trace the evolution
 from the molecular cloud density $\sim 100 {\rm cm}^{-3}$ to
 that of the protostellar disk $\sim 10^{10} {\rm cm}^{-3}$ or more.
Fluctuation with small amplitude grows by the gravitational instability.
It forms a disk perpendicular to the magnetic fields which runs in the
 direction parallel to the major axis of the cloud.
Matter accrets on to the disk mainly flowing along the magnetic fields and
 this makes the column density increase.
The radial inflow, whose velocity is slower than that perpendicular to the
disk,
 is driven by the increase of the gravity.
While the equation of state is isothermal and magnetic fields are
 perfectly coupled with the matter, which is realized in the density range of
 $\rho \lesssim 10^{10}{\rm cm}^{-3}$, never stops the contraction.
The structure of the contracting disk reaches that of a singular solution
 as the density and the column density obey
 $\rho(r)\propto r^{-2}$ and $\sigma(r)\propto r^{-1}$, respectively.
The magnetic field strength on the mid-plane is proportional to
 $\rho(r)^{1/2}$ and further that of the center ($B_c$) evolves
 as proportional to the square root of the gas density
 ($\propto \rho_c^{1/2}$).
It is shown that isothermal clouds experience ``run-away'' collapses.
The evolution after the equation of state becomes hard is also discussed.
\end{abstract}


\section{Introduction}

Molecular clouds are often observed as filaments.
One example is $\rho$ Ophiuchi cloud, in which
 two filaments extends from $\rho$ Oph main cloud ($l\simeq 353\deg$,
 $b\simeq$ 17\deg) and
 $\rho$ Oph East cloud ($l\simeq 354\deg$, $b\simeq 16\deg$),
 respectively.
The filaments have full lengths of 10-17.5 pc, although their
 widths are as narrow as 0.24 pc (Loren 1989).
In the Taurus molecular cloud (this cloud looks like a couple of filaments as a
 whole), between Heiles Cloud 2 and L1495,
 a filamentary dark cloud (B213) is seen.
Its size is approximately 20\arcmin $\times$ 3\deg $\simeq$ (1.2 pc
 $\times$ 11 pc) (Fukui \& Mizuno 1991).
The axial ratio reaches $\sim 10$ in this case, too.
Filaments are found even in giant molecular clouds
 (such as the Orion L1641 cloud: Bally 1989) which are known as
 sites of massive star formation.
In Figure  10 of Bally (1989), the topology of the magnetic field is shown
 for L1641, which indicates the magnetic fields in the cloud are running
 in the direction parallel to the major axis of the cloud, while just
 outside of L1641 toroidal magnetic fields are found lapping the filament.

Collapse and fragmentation in the cylindrical cloud are studied by
 Bastien and collaborators for non-magnetic clouds (Bastien 1983);
 Tomisaka (1995) and Nakamura, Hanawa, \& Nakano (1994) for magnetized clouds.
Summarizing result by Tomisaka (1995; hereafter referred to as Paper I)
 and Nakamura et al (1994),
 the process of fragmentation and collapse in the cylindrical magnetized
 isothermal cloud is as follows:
The eigen-mode which has the most unstable growth rate predominates even from
 random fluctuations with small amplitudes and the cylindrical cloud fragments
 into prolate spheroidal condensations separated by every $\lambda \simeq
 20 c_s/(4\pi G \rho_c)^{1/2}$ where $c_s$ and $\rho_c$ represent,
 respectively, the isothermal sound speed and the gas density on the
 symmetric axis of the cylinder.
This condensation contracts mainly in the direction parallel to the
 magnetic field (axis of the cylinder).
Finally forms a disk which runs perpendicularly to the magnetic field.
The disk, which continues to contract, has the density distribution similar to
 that of a singular solution for isothermal hydrostatic sphere
 (Chandrasekhar 1939):
 $\rho_{\rm sing} = c_s^2/2\pi G r^2$ in the $r$-direction.
In $z$-direction, the density distribution is much different from the above
 but the scale height in $z$-direction is much smaller than that of
 $r$-direction.
Accretion occurs mainly in the direction parallel to the magnetic field
 ($v_z \gg v_r$).

Since as the contraction proceeds
 the density in the center of the disk increases,
 spatial resolution is needed to see the evolution completely.
First method to resolve it is to increase the mesh number.
However, time necessary for the calculation increases at least proportional
 to $\propto$(mesh number in one dimension)$^3$ (see Appendix of Paper I).
Thus, it is difficult to use a grid with meshes in one dimension
 $N \gtrsim 1000$ even using massive supercomputers.
Here, we adopt the ``nested grid method'' (Ruffert 1992;
 Yorke, Bohdenheimer, \& Laughlin 1993).
In this scheme, several levels to grids are prepared:
 a fine grid which covers only the disk center,
 and a coarse grid which covers the whole cloud.
The grids system is overlaid with each other and  connected
 as the boundary conditions for a fine grid is determined by a
 coarse grid and contrarily physical quantities of a coarse grid
 are calculated by those of an overlaid fine grid.
This idea was originally developed by Berger \& Oliger (1984) and
Berger \& Colella (1989).

\section{Models and Numerical Method}

Since for the dense gas found in interstellar clouds with $\simeq 10$K,
 the equation of state is well approximated with the isothermal one
 (Larson 1985), we assume here that the gas obeys the isothermal
 equation of state.
While the gas density does not exceed $n\gtrsim 10^{10}{\rm cm}^{-3}$,
 the decouple process of the magnetic fields due to Joule loss
 is not effective (Nakano 1988).
Further the time scale of the ambipolar diffusion/plasma drift
 (the time scale that the neutral molecules flow toward the cloud
 center across the magnetic fields and thus charged ions) is estimated as
 $\sim 10$ times the free-fall time (Nakano 1988).
Since the time scale in which the gravitational instability grows is
 shorter than that of the ambipolar diffusion, it is reasonable
 if we assume the magnetic fields are perfectly frozen-in the matter
 even for the neutral matter.
Thus, we study the evolution of the cloud under the assumptions of
 isothermal equation of state and ideal magnetohydrodynamics (MHD).

As in Paper I, we assume the length of the cylindrical cloud
 is much longer than the width of it and we apply a periodic boundary
 condition in the direction of the axis of the cloud.
The length of the numerical box is a priori arbitrary.
Since it is shown, in Paper I, that only the wave with the maximum growth
 rate appears even from a white noise, we restrict ourselves to a
 numerical box whose size is equal to the most unstable wavelength
 known by the linear stability theory (e.g., Elmegreen \& Elmegreen 1978;
 Nagasawa 1987) as
\begin{equation}
\lambda_{\rm max} \simeq \frac{20\gamma c_s}{(4\pi G \rho_c)^{1/2}},
\label{lambda-max}
\end{equation}
where $\rho_c$ means the density on the symmetric axis of the cylinder
 and $\gamma\simeq 1$ represents a numerical factor obtained from numerical
 calculation (Nakamura, Hanawa, \& Nakano 1993).

\subsection{Initial Condition}

The initial condition is the same as Paper I.
That is, we assume that the isothermal cylindrical cloud which is
 in a magnetohydrostatic equilibrium and immersed in an
 external pressure, $p_{\rm ext}=c_s^2 \rho_s$, where $c_s$ and $\rho_s$
 represent the isothermal sound speed in the cloud and the gas density
 on the surface of the cloud.
To mimic the situation that fluctuations with small amplitudes
 grow with the gravitational instability, we add density inhomogeneity
 with a small amplitude.
When a ratio of the magnetic pressure $B_z^2/8\pi$ to the
 thermal one is assumed constant, $\alpha/2 \equiv B_z^2/8\pi/p=$
 constant\footnote{$\alpha^{1/2}$ is the parameter which represents the ratio
 of the Alfv\'{e}n speed and the isothermal sound speed.},
 the density distribution in hydrostatic balance is given by
\begin{equation}
 \rho_0(r)=\rho_c/{\left[1+\fracd{\pi G \rho_c r^2}
                         {2 c_s^2(1+\alpha/2)}\right]^2},
 \label{eqrho}
\end{equation}
where no toroidal component of the magnetic fields exists, $B_\phi=0$.
Scaling in a unit such as $4 \pi G = c_s = p_{\rm ext} = 1$,
 this is rewritten as
\begin{equation}
 \rho_0(r)=\rho_c/\left[1+\fracd{\rho_c}{8}\fracd{r^2}{1+\alpha/2}\right]^2,
 \label{eqFr}
\end{equation}
where the distance is normalized with $H=c_s/(4\pi G \rho_s)^{1/2}$,
and  the density is normalized with that on the surface of the cloud
 $\rho_s=p_{\rm ext}/c_s^2$.
The radial density distribution is seen in Figure 1 of Paper I.
Parameters used here are summarized in Table\ref{tbl-1}.
\begin{table}
\begin{center}
\begin{tabular}{lrrrrrr}
 Model &
 \multicolumn{1}{c}{$\alpha$} &
 \multicolumn{1}{c}{$F$} &
 \multicolumn{1}{c}{$\delta$} &
 \multicolumn{1}{c}{$N$} &
 \multicolumn{1}{c}{$l_z$} &
 \multicolumn{1}{c}{Eq.~of State} \\
\tableline
\tableline
A...... &   1 & 100 & $10^{-2}$ & 14 & 1.935 & Isothermal \\
B...... & 0.1 & 100 & $10^{-2}$ & 14 & 2.21  & Isothermal \\
C...... &   1 & 100 & $10^{-2}$ & 14 & 1.935 & Composite\tablenotemark{a} \\
D...... &   1 & 100 & $10^{-2}$ & 14 & 1.935 & Composite\tablenotemark{b} \\
\tableline
\end{tabular}
\end{center}

\tablenotetext{a}{Isothermal for $\rho < 10^8 \rho_s$
and $p \propto \rho^{4/3}$ for higher density.}

\tablenotetext{b}{The same as Model C but the critical density
 $\rho_{\rm crit}$ is taken as $10^{6}\rho_s$.}

\caption{Model Parameters.} \label{tbl-1}

\end{table}

As a fluctuation added to $\rho_0$, we assume a sinusoidal function
 with a wavelength $l_z$,
\begin{equation}
 \rho(z,r)=\rho_0(r)(1-\delta)\cos(2\pi z/l_z),
 \label{eqdel}
\end{equation}
where the relative amplitude $\delta$ is taken small enough
 ($\delta = 10^{-2}$).

\subsection{Basic Equations}

Basic equations are the unsteady MHD equation and the Poisson equation
 for the gravitational potential.
We assume no toroidal fields (neither magnetic fields nor velocity).
In the cylindrical coordinate ($z$, $r$, $\phi$)\footnote{($z$,$r$,$\phi$)
 is the right-handed coordinate system. Thus, the $z$-axis is illustrated
 as a horizontal axis in figures contrary to a common usage.}
 with $\partial / \partial \phi=0$, they are expressed as follows:
\begin{equation}
 \dif{\rho}{t}+\dif{}{z}(\rho v_z)+\frac{1}{r}\dif{}{r}(r\rho v_r)=0,
 \label{eqMHD1}
\end{equation}
\begin{eqnarray}
 \dif{\rho v_z}{t}+\dif{}{z}( \rho v_z v_z) +
  \frac{1}{r}\dif{}{r}(r\rho v_z v_r)= - c_s^2 \dif{\rho}{z}
     -\rho\dif{\psi}{z} \nonumber \\
     -\frac{1}{4\pi}\left(\dif{B_r}{z}-\dif{B_z}{r}\right)B_r,
 \label{eqMHD2}
\end{eqnarray}
\begin{eqnarray}
 \dif{\rho v_r}{t}+\dif{}{z}( \rho v_r v_z) +
  \frac{1}{r}\dif{}{r}(r\rho v_r v_r)= - c_s^2 \dif{\rho}{r}
     -\rho\dif{\psi}{r} \nonumber \\
     +\frac{1}{4\pi}\left(\dif{B_r}{z}-\dif{B_z}{r}\right)B_z,
 \label{eqMHD3}
\end{eqnarray}
\begin{equation}
  \dif{B_z}{t}=\frac{1}{r}\dif{}{r}[r(v_zB_r-v_rB_z)],
  \label{eqMHD4}
\end{equation}
\begin{equation}
  \dif{B_r}{t}=-\dif{}{z}(v_zB_r-v_rB_z),
  \label{eqMHD5}
\end{equation}
\begin{equation}
  \frac{\partial^2 \psi}{\partial z^2}
 + \frac{1}{r}\dif{}{r}\left(r\dif{\psi}{r}\right) =
   4 \pi G \rho,
  \label{eqPoi}
\end{equation}
 where the variables have their ordinary meanings.
Equation (\ref{eqMHD1}) is the continuity equation;
equations (\ref{eqMHD2}) \& (\ref{eqMHD3}) are the equations of motion.
The induction equations for the poloidal magnetic fields are
 equations (\ref{eqMHD4}) \& (\ref{eqMHD5}).
The last equation (\ref{eqPoi}) is the Poisson equation.

\subsection{Numerical Scheme}

As is seen in Paper I, the density distribution which is finally
 achieved in the gravitationally contracting disk is rather
 singular as $\rho(r) \simeq A r^{-2}$.
Therefore, near the center of the disk, the density increases
 monotonically and thus the density scale-height decreases.
Any numerical schemes do not guarantee the stabilities after the quantities
 differ much in adjacent cell-to-cell, although a specific value of
 the ratio over which the calculation becomes unstable depends upon
 a scheme.
In Paper I,
 a nonuniformly spaced grid was used in which the grid spacing in the
 $z$-direction increases as the distance from the disk midplane
 increases was used.
However, this leads to a mesh whose shape is far from the square
 ($\Delta z=\Delta r$), i.e.,
 $\Delta z \gg \Delta r$
  (near the pole $r=0$ but far from the disk center) or
 $\Delta z \ll \Delta r$
  (near the disk midplane but far from the $z$-axis).
Unnecessarily close spacing in the above has disadvantages since
 the numerical time step is severely limited there.
Further, it is thought that schemes using the square grids are
 more robust than those using rectangle grids
 ($\Delta z \ne \Delta r$).

In the nested grid scheme, all cells are square.
In this scheme, several levels of grid systems are prepared;
 a fine grid covers the disk center
 and a coarse grid is for a whole cloud.
Respective grids are overlaid as a co-centered fashion with each other
 as Figure 1.
Fifteen levels of grids are used in the present paper,
 which are named as L0 (the coarsest), L1, L2, $\ldots$, L14 (the finest).
Since the mesh spacing $\Delta z= \Delta r = \Delta h$ of the $n$-th level
 is just a half of that of the $(n-1)$-th level,
 the spacing of the L14 grid is equal to $1/2^{14}=1/16364$ of that of L0,
 that is, $\Delta h_{14}=\Delta h_0/16364$.
Thus the scheme has a wide dynamic range in the space dimension.
Since we take 64 meshes in one-dimension for each levels (L$n$ for $n=0,1,
 \ldots , 14$), the grid spacing in L14 corresponds to $1/1048576\simeq
 1/10^6$ of the size of the coarse grid L0 ($l_z$).

Boundary values for a fine grid (L$n$) are determined by interpolation of
 the values of a coarser grid (L$n-1$) and the physical quantities in a
 coarse cell (L$n-1$) are calculated by averaging those in four fine cells
 (L$n$) overlaid onto the coarse mesh.
Detailed procedure is written in Appendix.
The simulation simulation begins with five levels (L0, L1, L2, L3 \& L4).
Depth of the grid levels is increased when spatial resolution is necessary,
 upto L14.

Numerical methods applied to each levels of grids are the
 same as those of Paper I.
(In other words, the scheme adopted in Paper I is identical
 with the nested grid scheme consisting of L0 only.)
Unsteady MHD equations are solved using van Leer's monotonic
 interpolation (1977) and the constrained transport method by Evans \&
 Hawley (1988).
We adopt MILUCGS (modified incomplete LU decomposition preconditioned
 conjugate gradient squared method: Meijerink \& van der Vorst 1977;
 Gustafsson 1978).
We compared the numerical result by the nested grid scheme
 with that of Paper I.
While the maximum density is less than $\sim 10^{4.5} \rho_s$,
 the density distributions are identical except for regions
 near the boundaries.
After the maximum density exceeds $\sim 10^5 \rho_s$,
 numerical oscillation appeared near the center of the disk in Paper I.
However, the nested grid can resolve smaller structures than Paper I
 seen in the nest section.

\section{Results}

Model A in the present paper is identical to Model B of Paper I.
Figure 2 shows the structure of the cloud at $t=1.367$
 when $\rho_c=10^5 \rho_s$.
This corresponds to Figure 3 of Paper I, which is the final state
 that can be reached by a scheme using 400$\times$400 meshes
 without the nested grid technique.
Gas falls down along the magnetic field and forms a disk running
 perpendicularly to the magnetic fields (Fig.2{\em a}).
Magnetic field lines are squeezed by a dense gas disk and form a
 valley in the center.
Figure 2{\em b} shows the structure seen using the L2 grid,
 whose spatial resolution is 4 times finer than L0.
The disk whose shape is concaved is captured well in L2.
Although the center of the disk looks like a sphere in L0,
 this is due to a low resolution of L0.
The maximum density attained is also restricted in L0 as
 $\sim 10^4 \rho_s$, while that in L2 reaches $\gtrsim 10^{4.8}\rho_s$
 ($\rho_c \sim 10^5 \rho_s$ for L$n$ $n \ge 3$).
The scheme with evenly spaced meshes adopted in Paper I could not
 proceeds further.

In Figure 3, we compare
 the structure at $t=1.372$ when $\rho_c = 10^6 \rho_s$ ({\em a})
 and that at $t=1.373$ when $\rho_c = 10^7 \rho_s$ ({\em b}).
Since the density increases monotonically in a short time scale,
 it seems better to choose the central density $\rho_c$ to indicate
 the time evolution, instead of the time passing after the simulation begins.
Comparing ({\em a}) and ({\em b}),
 it is clearly seen that the density distribution only near the
 disk center evolves;
The gas accrets from the region near the axis flowing
 perpendicularly to the disk;
The evolution time scale for the disk far from the $z$-axis is much longer.

Figure 4{\em a} and {\em b} shows the disk covered in the L9 grid.
It is seen that the disk continues to contract in the very center.
Comparing the directions of velocities and magnetic fields,
 we can see the gas flows in the direction parallel to the magnetic
 field outside the disk (a region with $\rho \lesssim 10^5$).
This means that the gas flow is controlled by the magnetic field and
 the magnetic field lines are not squeezed strongly there.
To the contrary, in the disk,
 the gas seems to move across the magnetic field line;
 the gas squeezes and drags the magnetic field lines to the center.
It is to be noted that the magnetic field strengths in and out of the disk
 ($\rho \gtrsim 10^5$ or $\rho \lesssim 10^5$) differ slightly,
 although the gas densities differ in three orders of magnitude.
This is very similar to the flow realized in a collapsing
 rotating isothermal cloud (Norman, Wilson, \& Barton 1980).
That is, the gas accreting from the direction of angular momentum
 moves mainly along the lines of constant specific angular momentum;
 the gas contracting radially in the disk seems to move across these lines.

Comparing Figures 4{\em a} and {\em b}, it is shown that the thickness
 of the disk decreases according to the increase of $\rho_c$;
The half thickness of the disk  decreases from $5\times 10^{-4}$ ({\em a})
 to $1.5\times 10^{-4}$ ({\em b}), while the central density increases
 from $10^{8.3}$ to $10^{10}$.
When $\rho_c$ reaches $10^{10}$, although the resolution obtained by L9
 ({\em b}) is insufficient, the L11 grid resolves the vertical
 structure of the disk even in the central high-density region.

In Figure 5, the cross-cut view along the $z$-axis ({\em a}) and that
 of $r$-axis ({\em b}) are plotted.
{}From Figure 5{\em a}, it is clear that the cylindrical cloud is
 divided into a disk and an intra-disk.
A half-height of the disk decreases as from
 $h\sim 10^{-2}$ for $\rho_c \sim 10^5$ to
 $h\sim 10^{-4}$ for $\rho_c \sim 10^{10}$.
Gas accrets to the disk with supersonic speed $|v_z| > 1$.
As long as the gas density in the disk does not exceed $\rho_c\sim
 10^6$, the infalling gas is gradually decelerated and forms a disk.
However, after the disk is developed $\rho_c \gtrsim 10^7$,
 the gas is stopped abruptly on the disk surface where the gas
 is compressed and the density increases much.
Comparing $B_z(z,0)$ and $\rho(z,0)$, in the intra-disk region
 the magnetic field strength is almost proportional to the density,
 e.g., $B_z(z,0) \sim A \rho(z,0)$.
The ratio of the thermal pressure to the magnetic pressure, $\beta$,
 decreases from $\beta \sim 1$ at $z=\pm l_z/2$ to $\beta \sim 10^{-2.5}$
 on the surface of the disk (the values are for $\rho_c=10^9$).
This explains why the inflow in intra-disk region is controlled
 by the magnetic field and the gas flows along the magnetic fields.
The $\beta$ value begins to increase after the gas enters the disk
 and $\beta$ reaches again $\sim 7.5$ reaching the center.
Thus the magnetic fields are dragged by radial motion of the gas.

To the contrary, there is no discontinuity in the radial distributions
 ({\em b}).
The density increases in a systematic fashion.
There seems to exist an asymptotic density distribution and
 the radial density distribution reaches it.
It is well fitted as $\rho(0,r)\simeq 20 r^{-2.08}$.
The power law distribution $\propto r^{-2}$ is seen in a course of collapse
 of the isothermal rotating cloud (Norman et al 1980; Narita et al 1984)
 as well as the
 isothermal spherical collapse (Larson 1969; Penston 1969).
The disk is divided into two parts:
 a core which shows $\rho(0,r)\simeq$constant
 and an envelope which shows $\rho \propto r^{-2}$.
The evolution proceeds as a fashion that the core size $r_c$
 decreases monotonically and the region of envelope predominates.

The magnetic field strength seems to approach to a singular distribution as
 $B_z(0,r)\simeq 2 r^{-1.06}$.
It is to be noted that the derived power is nearly equal to -1 as
 $B_z(0,r) \propto r^{-1}$, which is a consequence of a tight
 relationship between $\rho(0,r)$
 and $B_z(0,r)$ as $B(0,r)\propto \rho(0,r)^{1/2}$
 (see Fig.5{\em c}; Tomisaka 1995; Nakamura et al 1994).
Since the magnetic fields are frozen-into the gas,
 the column density integrated along the magnetic field line is
 proportional to the magnetic field strength.
If the disk is approximated as an isothermal {\em plane-parallel} disk,
 the column density $\sigma$ is related to the midplane density $\rho_0$
 as
\begin{equation}
  \rho_0 = \frac{\pi G}{2c_s^2}\sigma^2+\rho_s,
  \label{eq.3-1}
\end{equation}
where $\rho_s$ means the surface density $=p_{\rm ext}/c_s^2$
(Spitzer 1978).
This indicates that $\rho_0 \propto \sigma^2$ when $\rho_0 \gg \rho_s$.
Thus, in the plane-parallel approximation, the magnetic field strength
 $B_z(0,r)$ is proportional to the square root of the density
 $\rho(0,z)$.

The column density which is integrated vertically as
 $\sigma = \int_{-l_z/2}^{+l_z/2} \rho dz$
is plotted in Figure 5{\em d}.
It shows that the distribution reaches a singular solution as
 $\sigma \propto r^{-1}$,
 although it shows a flat distribution in an early phase.
The singular power law distribution is approximated as
 $\sigma \simeq 15 r^{-1.02}$.
{}From Figure 5{\em d}, it is shown that the core size, $r_c$,
 which is defined as the region where $\sigma\simeq$ constant,
 decreases with time and the column density in the core $\sigma_c$
 is proportional to $1/r_c$, that is, $\sigma_c\times r_c \simeq$
 constant.
This is seen in a course of ``run-away collapse'' in rotating isothermal
 clouds (Norman et al 1980; Narita et al 1984).
In the run-away collapse, the density in a small part of the cloud
 increases infinitely, even the centrifugal force prevents the cloud
 from global contraction.
In contrast to the angular momentum, the magnetic field can not
 support a cloud as a whole when the mass is larger than the critical mass
 (Mouschovias \& Spitzer 1976; Tomisaka, Ikeuchi, Nakamura 1988, 1989).
Although the angular momentum and the magnetic fields have
 different effects on the global contraction,
 it is to be noted that the high density core in the disk cloud
 contracts as a `run-away collapse' in isothermal clouds.
As a result, it is shown that
 the radial distributions are reaching singular form as
\begin{equation}
 \rho(r) \propto B_z(r)^2 \propto \sigma(r)^2 \propto r^{-2}.
\end{equation}

The radial inflow velocity is supersonic after $\rho_c > 10^5$.
However, it is to be noticed that
 the radial velocity is much slower than the vertical velocity.
This is due to the effect of the magnetic fields;
 that is, the magnetic pressure is comparable with the thermal pressure
 in the disk and thus the motion across the magnetic fields is more or
 less blocked.

\subsection{Weak Magnetic Fields}

Model B corresponds to a case with weak magnetic fields ($\alpha = 0.1$).
Since the magnetic fields support the cloud in the $r$-direction,
 the radial size of the cloud is reduced as $R \simeq 0.8694$ compared with
 $R \simeq 1.039$ of Model A.
The structures for three stages are shown in Figure 6.
The difference from Model A is apparent if compared Figure 6{\em a}
 ($t=1.612$) with Figure 2{\em a} ($t=1.367$),
 which has the similar $\rho_c$.
At that epoch, the shape of a region which has a positive density
 fluctuation ($\rho > 10^2$) is a prolate spheroid, which coincides with
 the eigen-function derived by the linear perturbation theory.
A disk is not formed yet but a spherical core is formed.
It has been shown in Paper I that in a {\sl non-magnetized} cylindrical cloud
 a spherical core is formed and it continues to collapse.
If the cloud contracts only in the $r$-direction, the magnetic pressure
 increases in proportional to the square of the thermal pressure.
Even an initial magnetic pressure is as small as 0.05$\times$
 thermal pressure,
 it becomes comparable to the thermal pressure when the central
 density exceeds $10^5$.
Thus before the epoch when $\rho_c \sim 10^5$, the magnetic field only has
 a subsidiary role in the dynamics of the cloud.

In Figure 6{\em b}, the next stage is shown.
The central density reaches $\gtrsim 10^7$ at that time and
 the magnetic pressure has an important role compared with the previous
 stage.
It is seen that
 a small disk is formed which runs perpendicular to the magnetic fields.
Comparing this with Figure 3{\em b}, both of which
 have similar central densities round $\sim 10^7$,
 the difference in the structures of the formed disks is clear.
It is evident that the lateral size of the disk becomes much
 smaller in case of weak magnetic fields.
A substantial part of the high-density region still forms a sphere
 even after a small disk contracts and the density reaches
 $\sim 10^{10}$ (Fig.6{\em c}).

The cross-cut views along $z$- and $r$-axis are shown in Figure 7.
The respective lines correspond to different epochs.
The characteristic points are similar to Model A;
 magnetic field strength $B_z(0,r)$ is proportional to the square root
 of the gas density.
It is shown that the distributions of density and magnetic fields
 reach asymptotically as
 $\rho(0,r) \simeq 2.5 r^{-2.4}$ and $B_Z(0,r) \simeq 0.32 r^{-1.26}$.
The column density distribution is well fitted by a power law
 singular solution as $\sigma = 24 r^{-0.974}$.

In Figure 8, the plasma $\beta$,
 which is the ratio of the thermal pressure to magnetic pressure,
 is plotted versus the distance from the center of the disk.
Figures 8{\em a} and {\em b} correspond to cross-cuts
 along the $z$-axis for Model A and B, respectively.
With reaching the center $\beta$ decreases and just outside the disk
 $\beta \lesssim 10^{-2}$ when $\rho_c \gtrsim 10^8$ for Model A.
The minimum of $\beta$ is affected by the flux contained in the cloud;
 in Model B it is $\beta \sim 0.1$ at an epoch of $\rho_c=10^{10}$.
In the disk, $\beta$ increases with reaching the center.
In the intra-disk region, the thermal and the magnetic pressures are
 well correlated as $\partial \log p / \partial \log p_{\rm mag} \simeq$
 0.6 (for Model A) and 0.75 (for Model B).

Figures 8{\em c} and {\em d} correspond to cross-cuts
 along the $r$-axis.
In contrast to $\beta(z,0)$, $\beta(0,r)$ increases with time in the disk.
For example, in the range of $r \gtrsim 10^{-4}$ the $\beta$ increases
 with time.
Noting that $\alpha=1$ is equivalent to $\beta=2$, it is clear
 that the outer part does not experience violent contraction.
In Model A, the $\beta$ seems to reach asymptotically $\simeq 10$.
In Model B this is found in the range of 15 -- 35.
Since $\beta$ decreases when the magnetic fields are compressed
 by the radial contraction but increases when the cloud contracts
 along the magnetic fields, the steady increase of $\beta$ indicates
 that disk formation continues.

Since the central column density $\sigma_c(r\simeq 0)$ is proportional to
 the magnetic field
 strength $B_z$, it is estimated as
\begin{eqnarray}
\sigma_c (t) & \simeq & \rho_{c~{\rm init}} \lambda_{\rm max}
                       (B_z/B_{z~{\rm init}}),\\
             & \simeq & \frac{20 \gamma c_s F^{1/2} \rho_s^{1/2}}
                             {(4\pi G)^{1/2}}
                       (B_z/B_{z~{\rm init}}),
\end{eqnarray}
where we used equation (\ref{lambda-max}) and the initial condition
(the suffix ``init'' represents the initial value).
Using equation (\ref{eq.3-1}), the plasma $\beta$ in the central region of
 the cloud is estimated as
\begin{eqnarray}
 \beta_c &\simeq& 50 \gamma^2 \beta_{c~{\rm init}},\\
         &=&100 \gamma^2 /\alpha.
\end{eqnarray}
This expects $\beta_c\simeq 100 - 1000$ for Models A and B.
However, the actual $\beta_c\gtrsim 10$ is much smaller than that expected
 from a thin plane-parallel disk model $\beta_c \sim 100- 1000$.
This suggests the gravity $g_z$ may be overestimated in the plane-parallel
 disk model.
Since the disk has a finite radial size, in other word, the column
 density $\sigma(r)$ decreases proportionally to $\propto r^{-1}$,
 the gravity should be weaker than that expected by a perfectly plane-parallel
 disk.

In the isothermal plane-parallel disk, the density distribution
 is expressed as
\begin{equation}
 \rho(z) = \rho_c\ %
       {\rm sech}^2 \left( \frac{z}{c_s/\sqrt{2\pi G \rho_c}} \right),
 \label{eq.sech2}
\end{equation}
where $\rho_c$ represents the density on the mid-plane $z=0$ (Spitzer 1978).
Comparing this with $\rho(z,0)$ of Figure 5{\em a},
 it is seen that the size is {\em underestimated}
 if we use the plane-parallel disk approximation:
When the size is measured by the $z$-coordinate where $\rho(z,0)=10^9$,
 equation (\ref{eq.sech2}) gives the size of $\simeq 2/3$ times smaller
 than the actual size in Figure 5{\em a} of $\rho_c=10^{10}$.
If we compare the size of $\rho(z,0)=10^8$, the factor is equal to
 $\simeq 0.4$.
This indicates that the gravity in the plane-parallel disk with
 an identical central density is stronger than that realized
 in the contracting disk here.

\section{Discussion}

\subsection{Effect of the Equation of State}

In the previous section, it is shown that the cylindrical cloud
 forms disks and the central part of the disk experiences
 the ``run-away collapse''.
However, this seems to due to the assumption of the isothermal equation
 of state for the gas.
Actually, before the gas density increases infinitely,
 the cloud core becomes optically thick and the radiation, which
 keeps the gas isothermal, begins to be trapped in the core
 (Hayashi 1966).
The critical density, above which the equation of state is no more
 isothermal, is estimated $\rho_{\rm crit}\sim 10^{-14}-10^{-12.5}
 {\rm g\,cm}^{-3}$ for the mass range of $100 M_{\sun} - 0.01 M_{\sun}$.
The gas obeys another equation state above $\rho_{\rm crit}$, which
 should be solved by the radiative magnetohydrodynamics.
Here, for simplicity, we assume the polytropic relation between the pressure
 and the density as
\begin{equation}
p = \left\{ \begin{array}{ll}
              c_s^2 \rho & \mbox{if $\rho < \rho_{\rm crit}$,} \\
              c_s^2 \rho_{\rm crit}
                          (\rho /\rho_{\rm crit})^{1+1/{\cal N}}
                            & \mbox{if $\rho > \rho_{\rm crit}$,}
            \end{array}
    \right.
\end{equation}
where $\cal N$ means the polytropic index.
In Model C we assume ${\cal N}=3$ and $\rho_{\rm crit}=10^8 \rho_s$.
If we adopt $\rho_s=100 {\rm cm}^{-3}$, this means $\rho_{\rm crit}=
 10^{10}{\rm cm}^{-3}$.

\subsubsection{Model C}

Since the difference is expected in the evolution
 after the maximum density
 is larger than $\rho_{\rm crit}$, we will focus on the
 central part of the disk.
Figure 9{\em a-c} shows the structure of the central disk.
It is noticed that there exists a steep density jump near
 $z\sim 2\times 10^{-4}$ and
 the accreting flow in the $z$-direction is stopped here.
The position of density jump seems stationary after the
 central density reaches $\rho_c \simeq 10^9$,
 while the shock front recedes continuously in isothermal
 models (Compare Fig.5{\em a} with  Fig.9{\em a}).
In contrast to the cross-cut view along the $z$-axis,
 there is a slight change in the radial distribution of density.
However, it is shown that the magnetic fields $B_z$
 is compressed by the effect of the thermal pressure,
 which is relatively higher than Model A
 (Fig.5{\em c} \& Fig.9{\em a}).
The plasma $\beta$ at the center of the disk decreases
 from $\simeq 6$ of Model A to $\simeq 3$.

Another difference is the central {\em spherical} core
 with a radius of $r\simeq 1.5\times 10^{-4}$.
This spherical core keeps its size.
In the case of isothermal clouds (Models A \& B),
 the accumulated mass results in the strong gravitational field
 as well as the pressure force.
To the contrary, the pressure force becomes relatively important
 over the gravity due to the harder equation state in this case.
Thus, the thermal pressure, which is isotropic in nature,
 plays an important role in the core and forms a spherical core.
If we assume the polytropic index $\cal N$ smaller than 3,
 (that is, a harder equation of state),
 we may find a bounce due to the thermal pressure
 similar to the cases of the spherical cloud (Larson 1969)
 and the rotating cloud (Narita et al. 1986).

\subsubsection{Model D}

The structure in the spherical core with the polytropic index
 ${\cal N}=3$ is simple for Model C as long as $\rho_c \lesssim
 100 \rho_{\rm crit}$.
To see the core evolution further,
 a model with $\rho_{\rm crit}=10^6$ is studied.
The evolution is traced in the range of 4 orders of magnitude of the
 central density, $10^6 \lesssim \rho_c \lesssim 10^{10}$.
The polytropic index is unchanged.

Figure 10 shows the structure of the central
 $|z| \lesssim 2\times 10^{-3}$ ({\em a}) and
 $|z| \lesssim 4\times 10^{-4}$ ({\em b}).
Two shock front systems are found:
 one is seen $|z| \sim 10^{-3}$ near the $z$-axis.
The other is $|z| \sim 1.5 \times 10^{-4}$ near the $z$-axis,
 which seems to be connected the density jumps seen in
 $r\simeq 4\times 10^{-4}$, $|z| \lesssim 2.5 \times 10^{-4}$.
The first shock is related to the density jump appeared in the
 previous model.
The structures for $\rho_c=10^6$, $10^7$, ..., $10^{10}$ are shown
 in Figure 10{\em c} and {\em d}.
It shows that a jump in $v_z$ appears after $\rho_c > 10^6$.
At first, the infalling velocity $v_z$ is completely stopped through
 the shock front and the core seems static ($\rho_c \sim 10^7$).
However, after $\rho_c \gtrsim 10^8$, the core begins to contract
 and the inflowing velocity in the core is accelerated.
Finally, another density/velocity jump appears when $\rho_c \sim 10^{10}$.
The second structure is seen in Figure 10{\em b} as
 a region where the density gradient is steep.

Oblique shocks are seen near $r\simeq 4\times 10^{-4}$.
Matter flowing from upper-right and upper-left directions
 ($|v_z|\sim |v_r|$) changes its direction through this
 surface to the radial direction ($|v_r| \gg |z_z|$).
Plotting the magnetic fields together, it changes its direction
 as the same sense as the velocity field.

It is concluded that the collapse in the center of the
 cloud/disk continues even after the equation of state
 becomes effectively harder than the isothermal one.
As the contraction proceeds, the infall velocity is accelerated
 and the second discontinuity is formed.
Discontinuity facing to the radial direction is also formed.
As a course of the non-isothermal contraction,
 spherical or quasi-spherical core is formed.
This shows that the smooth ``run-away'' collapse seems to be
 seen in the {\em isothermal} collapse and in the actual
 interstellar environment, shocks are inevitably formed
 in the dense core.

\subsection{Evolution of a Cylindrical Cloud}

It is shown that the cylindrical cloud threaded by magnetic fields
 running in the direction parallel to the symmetric axis
 is fragmented by the gravitational instability (Paper I and the
 present paper).
The fragment which takes initially prolate spheroidal shape
 contracts along the magnetic field and forms a disk.
Finally, the central part of the disk begins ``run-away collapse''.
Is this a common evolution track?

To answer the question, we have to start with the result of the
 linear instability theory.
Nakamura et al. (1993) obtained eigen-functions and growth rates
 of the perturbation added to the isothermal cylinder.
The wave-length which has the maximum growth rate is approximated
 as $\simeq 20 \gamma c_s/\sqrt{4\pi G \rho_c}$.
The fact that the expression does not contain the magnetic field
 strength is important.
This means that even when the cylindrical cloud is supported
 in the radial direction
 by the magnetic field, the cloud fragments with a characteristic
 size of scale-length determined by only the isothermal sound
 speed, $c_s$, and the central density, $\rho_c$.
The fragmentation process is driven by the Jean
 instability and the Jean mode is activated by the motion in
 the $z$-direction.
However, the magnetic fields have no effect to the motion
 parallel to it.
This is the reason why the maximum growth wave-length
 is not related to the magnetic fields.
(Precisely speaking, the maximum growth wave-length is dependent on
 $\alpha$ as $\lambda_{\rm max}\propto \alpha^{1/6}$ for $\alpha \gg 1$.
 However its dependence is weak [see Nakamura et al 1993].)

As shown in Paper I, if the fragmented disk has enough mass-to-flux
 ratio, $M/\Phi=\sigma/B_z$, the resultant disk is supercritical and
 must contract as a whole.
The critical value of the mass-to-flux ratio is $M/(\Phi/ G^{1/2}) \simeq
 0.17 \simeq 1/(2\pi)$ (Mouschovias \& Spitzer 1976; Tomisaka,
 et al 1988, 1989).
Since the mass-to-flux ratio for the disk formed in the cylindrical cloud
 is estimated as
\begin{equation}
 \frac{M}{\Phi/G^{1/2}} \simeq \gamma \frac{1.59}{\alpha^{1/2}}
\end{equation}
 (equation[4.4] of Paper I),
models that we calculated in the present paper lead to supercritical
 disks.
It is reasonable that the ``run-away'' collapse is seen in all models.

How about a cloud with strong magnetic fields?
Since $\alpha$ increases as the cloud contracts radially,
 it is realized as a consequence of contraction.
When hydrostatic balance is achieved after contraction,
 the gravitational instability grows.
If the value of $\alpha$ at that time is over $\sim 100$,
 the mass-to-flux ratio of the first growing fragment may
 be smaller than the critical value.
However, it should be noted that
 the fragment which is more massive than the most unstable perturbation
 {\em can grow} even with a smaller growth rate than the most unstable one.
This means the fragments tend to merge each other
 moving parallelly to the magnetic fields.
This makes the mass increase but keeps the magnetic flux unchanged.
As a result, even when the cloud is compressed and
 $\alpha$ in the central part of the cloud is over 100,
 it is unlikely that many disks are formed and exist stably.

\subsection{Application to Dimensional Values}

The physical values were scaled with units of $c_s=4 \pi G = p_{\rm ext}=1$.
In this subsection, let us convert the non-dimensional values
 to dimensional ones.
Assuming the temperature of $T=10$K and molecular weight
 of $\mu=2.33$,
 the isothermal sound speed for this temperature is
 $c_s=190 {\rm m\,s}^{-1}(T/10{\rm K})^{1/2}$.
This leads to length scale of
\begin{eqnarray}
 H &   =    & c_s/\sqrt{4\pi G \rho_s}, \nonumber \\
   & \simeq & 0.36 {\rm pc}
             \left( \frac{n_s}{100{\rm cm}^{-3}}\right)^{-1/2}
             \left( \frac{c_s}{200 {\rm m\,s}^{-1}} \right),
\end{eqnarray}
where $n_s$ means the particle density on the surface of the cloud.
Since the radii of the initial states  of Models A and B
 are equal to $R=0.8694$ and 1.039 respectively,
 the radial size of the cloud corresponds to
 $\sim 0.4 {\rm pc}(n_s/100{\rm cm}^{-3})^{-1/2}(c_s/200{\rm m\,s}^{-1})$.
The initial cloud has the density contrast of $F=\rho_c/\rho_s$.
Typical densities in the cloud center varies from
 $n_c=10^4{\rm cm}^{-3}(F/100)(n_s/100{\rm cm}^{-3}) $
 to $10^{12}{\rm cm}^{-3}$ $(\rho_c/10^{10}\rho_s)$
 $(n_s/100{\rm cm}^{-3})$.
The magnetic field is normalized by a unit of
\begin{eqnarray}
 B_0 &   =    & \sqrt{4\pi \rho_s c_s^2}, \nonumber \\
     & \simeq & 1.4 ~ \mu {\rm G}
       \left( \frac{n_s}{100{\rm cm}^{-3}} \right)^{1/2}
       \left(\frac{c_s}{200{\rm m\,s}^{-1}}\right).
\end{eqnarray}
Time is measured with a unit of
\begin{eqnarray}
 \tau &   =    & 1/\sqrt{4\pi G \rho_s}, \nonumber \\
      & \simeq & 1.75 ~ {\rm Myr}
           \left( \frac{n_s}{100{\rm cm}^{-3}} \right)^{-1/2}.
\end{eqnarray}
Since perturbations in the cylindrical cloud grow
 and form disks in a time scale of $t\simeq 1.3-1.6$ for Models
 A and B,
 the physical time scale for the cylindrical cloud to experience
 the run-away collapse is estimated as
 $\sim (2-3){\rm Myr} ({n_s}/{100{\rm cm}^{-3}})^{-1/2}$.

The separation between the disks is estimated
\begin{eqnarray}
l_z &   =  & 20\gamma F^{-1/2} H, \nonumber \\
    & \sim & 0.72 ~ {\rm pc} ~ \gamma \left(\frac{F}{100}\right)^{-1/2}
      \left(\frac{n_s}{100{\rm cm}^{-3}}\right)^{-1/2} \nonumber \\
    & &   \left(\frac{c_s}{200{\rm m\,s}^{-1}}\right).
\end{eqnarray}
This indicates that the disk separation is
 approximately equal to the diameter of the cylindrical cloud.
Since the line density of the cloud is equal to
\begin{eqnarray}
 \lambda &    =   &\int_0^R 2\pi\rho r dr, \nonumber \\
         & \simeq & (2c_s^2/G)(F^{1/2}-1)/F^{1/2}(1+\alpha/2), \nonumber \\
         & \simeq & 25 ~ M_{\sun}{\rm pc}^{-1}
         \left( \frac{c_s}{200{\rm m\,s}^{-1}}\right)^2,
\end{eqnarray}
for  $F=100$ and $\alpha=1$,
the mass of the resultant disk is estimated as
\begin{eqnarray}
 M &    =   & l_z \lambda, \nonumber \\
   & \simeq & 18 ~ M_{\sun} \gamma
         \left( \frac{c_s}{200{\rm m\,s}^{-1}}\right)^3
         \left( \frac{n_s}{100{\rm cm}^{-3}}\right)^{-1/2}.
\end{eqnarray}
Singular power law distributions of
 the density ($\rho(0,r)\simeq 20 r^{-2.08}$),
 the magnetic field ($B_z \simeq 2 r^{-1.06}$),
 and the column density ($\sigma \simeq 15 r^{-1.02}$)
 for Model A
 are written as follows:
\begin{eqnarray}
 \rho(r) &\simeq& 2 \times 10^9 ~ {\rm cm}^{-3}
        \left(\frac{r}{100 {\rm AU}}\right)^{-2.08}\nonumber \\
        & &\left( \frac{n_s}{100{\rm cm}^{-3}}\right)
        \left( \frac{H}{0.36 {\rm pc}}\right)^{2.08},\\
 B_z(r) &\simeq& 3 ~ {\rm mG}
        \left(\frac{r}{100 {\rm AU}}\right)^{-1.06}
        \left( \frac{c_s}{200{\rm m\,s}^{-1}}\right)\nonumber\\
        & & \left( \frac{n_s}{100{\rm cm}^{-3}}\right)^{1/2}
        \left( \frac{H}{0.36 {\rm pc}}\right)^{1.06}, \\
 \sigma(r) &\simeq& 5 ~ {\rm g\,cm}^{-2}
        \left(\frac{r}{100 {\rm AU}}\right)^{-1.02}
        \left( \frac{c_s}{200{\rm m\,s}^{-1}}\right)\nonumber\\
        & & \left( \frac{n_s}{100{\rm cm}^{-3}}\right)^{1/2}
        \left( \frac{H}{0.36 {\rm pc}}\right)^{1.02}.
\end{eqnarray}

Recently, it is found that a gas disk around HL Tauri is contracting
 towards the center (Hayashi, Ohashi, \& Miyama 1993).
The disk with a mass of $M_{\rm disk} \sim 0.022 - 0.11 M_{\sun}$
 in $R=1400$ AU shows an indication of inflow
 with a speed of $v_r \sim 1 ~ {\rm km\,s}^{-1}$.
If this object is a disk which is in a stage of the run-away collase
 we studied, the observed contraction speed inevitablly leads to
 a large sound speed  $c_s \sim 0.5 {\rm km\,s}^{-1}$,
 because the radial inflow speed is approximately equal to $\sim 2 c_s$.
But, since this gives a relatively large disk mass as
 $M_{\rm disk} \sim \int_0^R \sigma 2\pi r dr \sim
 1 M_{\sun} (c_s/0.5{\rm km\,s}^{-1})(n_s/100{\rm cm}^{-3})^{1/2}
(H/0.36{\rm pc})^{1.02}$ in $R=$1400 AU,
 HL Tauri seems an object in a free-fall stage or the inside-out
 collapse stage (Galli \& Shu 1993a, b) subsequent to the run-away collapse.
Thus, if objects are found with relatively low contraction speed
 $v_r \sim 2 c_s$, this might be an object just before the
 run-away collapse.

\acknowledgments

This work was supported in part by the Grants-in-Aid from the Ministry
 of Education Culture and Science (05217208) in fiscal years 1993-1994.
Numerical calculations were done in part by supercomputer:
 S-3800/480 in University of Tokyo and VPP500/7 in ISAS.

\appendix
\section{Nested Grid Scheme}

In Appendix, the nested grid scheme is described very briefly.
We used 15 levels of the grids, in which finer grids are prepared
 for the very center of the contracting disk and coarser grids are
 to calculate the whole cloud.
As seen in Figure 1, a finer grid is wholly covered by a coarser grid.
We name respective grids as L0, L1, $\ldots$, L14.
L0 is the coarsest grid and L14 is the finest.
In the scheme, we choose a mesh size of a finer grid (L$n$)
 as 1/2 of that of the coarser grid (L$n-1$).
Since the mesh number of the respective grids are taken the same (64),
 the size of the numerical box of the coarser grid (L$n-1$)
 is twice as large as that of the finer one (L$n$).
For the character of the problem, all multiple grids are put in a co-centered
 fashion; the coordinates of the centers of the $z$-axis are taken
 identical for all grids (Fig.1).

Basic equations are the unsteady MHD equation and the Poisson equation
 (eq. \ref{eqMHD1} - \ref{eqPoi}).
As a boundary condition for L$n$, quantities on the upper ($z=Z_{\rm u}$),
 lower ($z=Z_{\rm l}$),
 and outer ($r=R_{\rm o}$) boundaries, such as $\rho$, {\bf v}, {\bf B}, and
 $\psi$,  are calculated from those of L$n-1$.
To derive the boundary values, we interpolate from those on coarser grids
 with van Leer's monotonic interpolation (van Leer 1977).
After determining the boundary values, we can solve the MHD and Poisson
 equations for L$n$.
Then, equations for L$n-1$ are solved.
Since one quarter of the mesh points of L$n-1$ are covered by the L$n$ grids,
 quantities of L$n-1$ in the overlapped region are overwritten by
 new values of L$n$.
To derive quantities of L$n-1$ we simply take the volume-average of the
 quantities in corresponding 4 cells of L$n$.
Quantities in the region overlapped by L$n-1$ are calculated
 but are not actually used, when equations for L$n-1$ are solved.
This makes data structure simple and the code achieves high
 performances on vector machines.

We show the algorithm here for three grids:
 coarse (L0), middle (L1) and a fine grid (L2), for simplicity.
\begin{itemize}
\setlength{\itemsep}{-2pt}
\item[(0.0)] Calculate boundary values for L1 from L0
  \begin{itemize}
  \setlength{\itemsep}{-2pt}
  \item[(1.0)] Calculate boundary values for L2 from L1
    \begin{itemize}
    \setlength{\itemsep}{-2pt}
    \item[(2.1)] Time step calculated from Courant condition ($dt21$)
    \item[(2.2)] Solve MHD eq. for L2
    \item[(2.3)] Solve the Poisson eq. for L2
    \end{itemize}
    \begin{itemize}
    \setlength{\itemsep}{-2pt}
    \item[(2.1)] Time step calculated from Courant condition ($dt22$)
    \item[(2.2)] Solve MHD eq. for L2
    \item[(2.3)] Solve the Poisson eq. for L2
    \end{itemize}
  \item[(1.1)] Time step for L1 is set as ($dt11=dt21+dt22$)
  \item[(1.2)] Solve MHD eq. for L1
  \item[(1.3)] Solve the Poisson eq. for L1
  \item[(1.4)] A part of L1 overlaid by L2 are replaced by quantities of L2
  \end{itemize}
  \begin{itemize}
  \item[(1.0)] Calculate boundary values for L2 from L1
    \begin{itemize}
    \setlength{\itemsep}{-2pt}
    \item[(2.1)] Time step calculated from Courant condition ($dt21$)
    \item[(2.2)] Solve MHD eq. for L2
    \item[(2.3)] Solve the Poisson eq. for L2
    \end{itemize}
    \begin{itemize}
    \setlength{\itemsep}{-2pt}
    \item[(2.1)] Time step calculated from Courant condition ($dt22$)
    \item[(2.2)] Solve MHD eq. for L2
    \item[(2.3)] Solve the Poisson eq. for L2
    \end{itemize}
  \item[(1.1)] Time step for L1 is set as ($dt12=dt21+dt22$)
  \item[(1.2)] Solve MHD eq. for L1
  \item[(1.3)] Solve the Poisson eq. for L1
  \item[(1.4)] A part of L0 overlaid by L1 are replaced by quantities of L1
  \end{itemize}
\item[(0.1)] Time step is set as $dt0=dt11+dt12$
\item[(0.2)] Solve MHD eq. for L0
\item[(0.3)] Solve the Poisson eq. for L0
\item[(0.4)] A part of L0 overlaid by L1 are replaced by quantities of L1
\end{itemize}

As is seen, in the $n$-th level, we calculate the MHD and Poisson equation
 of L$n+1$ twice and then go to L$n$.
The time step of L$n$ is equal to the total of two time steps of L$n+1$.
In the present paper, the time step is determined by the finest grid L$N$
 and used in coarser grids L0, L1, $\ldots$, L$N-1$.
The finest grid level are changed with necessity; that is,
 at first $N$ is equal to 4; and a finer grid level is generated
 ($N\rightarrow N+1$)
 when a density contrast in the finest grid becomes larger than 10.

We solved the Poisson equation by the ``modified incomplete LU decomposition
 preconditioned conjugate gradient squared method (MILUCGS: Meijerink \&
 van der Vorst 1977; Gustafsson 1978).
For MHD equations we use ``monotonic scheme'' (van Leer 1977,
 Norman \& Winkler 1986) and ``constrained transport scheme''
 (Evans \& Hawley 1988).
Routines to solve MHD equations (0.2, 1.2, \& 2.2)
 in the above are actually an identical subroutine
 as well as the Poisson solver (0.3, 1.3, \& 2.3) are identical
 with one another.
Since the mesh numbers of each levels are taken as the same,
 we can use the identical routine for data belonging to
 different levels of grids.
Due to this simplicity, the finest grid level ($N$) is changed easily
 without changing the code.
The code for a single level of grids is the same as the previous paper
 (Tomisaka 1995).


\onecolumn

{\bf Fig.1.} --- Explanation of the nested grid method.
Level 0 covers the entire cloud.
Level 1 covers the central 1/4 of L0,
Level 2 covers the central 1/4 of L1, and so on.
Since the numbers of the meshes of respective levels
 are the same (64$\times$64), the L1 has twice higher resolution
 than L0.

{\bf Fig.2.} ---
Snapshots of Model A at the epoch of $\rho_c=10^5$ ($t=1.367$).
The {\em Left} panel ({\em a}) shows the structure in L0 (the coarsest grid)
 and the {\em right} panel ({\em b}) shows that of L2
 (four times closer than L0).
Density contour lines (thick lines) and magnetic field lines (thin lines)
 are illustrated.
The contour levels of the denisties are shown by numbers attached to the
 contour lines as $\log_{10}\rho$.
Numbers seen on the right-bottom corner represent the maximum and
 minimum values of $\log \rho$ appeared in respective panels.

{\bf Fig.3.} --- Two snapshots of Model A:
 the {\em left} panel corresponds to the epoch of $\rho_c=10^6$
 ($t=1.372$) ({\em a})
 and the {\em right} shows that of $\rho_c=10^{7.3}$ ($t=1.373$) ({\em b}).
The fifth level of  grids L5 are plotted.
Density contours and velocity fields are illustrated.
Velocity vectors are illustrated every 3$\times$3 meshes.
Numbers seen above the density range means the maximum value
 of $|{\bf v}|$ (3.36 left and 4.02 right).

{Fig.4.} --- The {\em upper} panels: L9 structures at the epochs of
 $\rho_c=10^{8.33}$ (left: {\em a}) and
 $\rho_c=10^{10.13}$ (right: {\em b}) for Model A.
The {\em lower} shows the state when $\rho_c=10^{10}$
 but in the L11 grid ({\em c}).
Density contour lines (thick lines), magnetic field lines (thin lines),
 and velocity fields (vectors) are illustrated.

{\bf Fig.4.} --- ({\em a}): Cross-cut view on the $z$-axis for Model A.
 $\rho(z,0)$ (solid lines), $B_z(z,0)$ (dashed lines),
  and $|v_z(z,0)|$ (dotted lines) are illustrated.
 Eight different epochs are snapshoted.
 $\rho_c=10^{3.01}$, $10^4$, $10^{5.05}$, $10^{6.01}$,
 $10^{7.3}$, $10^{8.33}$, $10^{9.32}$ and $10^{10.13}$
  are plotted.
({\em b}): Cross-cut view on the $r$-axis.
 $\rho(0,r)$ (solid lines), $B_z(0,r)$ (dashed lines),
  and $-v_r(0,r)$ (dotted lines) are shown.
 The epochs of the snapshots are the same as {\em a}.
({\em c}): The correlation between density and magnetic field
  strength on the disk midplane $z=0$.
 Eight lines of $\rho(0,r)$-$B_z(0,r)$ correspond to the
  respective epochs of {\em a}.
 The left-bottum corner is the surface of the cloud and the
  right-upper ends are the disk centers.
({\em d}): The column density distribution against $r$.
 $\sigma(r)=\int \rho dz$.
 The epochs of the snapshots are the same as {\em a}.

{\bf Fig.6.} ---
The structure of the fragment in the cylindrical cloud for Model B.
({\em a}): the snapshot at $t=1.612$ ($\rho_c=10^{5.17}$) in L2.
 High-density part of the cloud indicates a prolate spheroidal shape
  and it contains a spherical core.
({\em b}): the snapshot at $t=1.616$ ($\rho_c=10^{7.17}$) in L5.
 A small disk is formed in the center of the spherical core.
({\em c}): the snapshot at $t=1.616$ ($\rho_c=10^{10.01}$) in L11.


{\bf Fig.7.} --- The same as Fig. 5 but for Model B.
Eight different epochs are snapshoted:
 $\rho_c=10^{3.03}$, $10^{4.09}$, $10^{5.17}$, $10^{6.5}$,
 $10^{7.17}$, $10^{8.38}$, $10^{9.19}$, and $10^{10.01}$.

{\bf Fig.8.} ---
The plasma $\beta$ versus the distance from the center of the disk
 $|z-z_c|$ for Models A ({\em a}) \& B ({\em b}).
Respective curves correspond to different epochs.
The epochs of the snapshots are the same as Figs.5 \& 7.
As the contraction proceeds the minimum of the $\beta$ decreases.
The plasma $\beta$ versus radial distance $r$
 for Models A ({\em c}) \& B ({\em d}).
Generally, the $\beta$ in the disk increases with time.
It indicates that the mass accretion continues on to the disk.

{\bf Fig.9.} --- Model C, in which the equation of state is assumed
 hard for $\rho > 10^8 \rho_s$.
This mimics the situation that the gravitational energy liberated
 in a course of collapse cannot be transferred after the
 cloud becomes optically thick.
({\em a}): snapshot when $\rho_c=10^{10.06}$. Density contour lines,
 magnetic fields, and velocity fields are plotted.
({\em b}) and ({\em c}): the cross-cut views along the $z$- and $r$-axis,
 when $\rho_c=10^{8.3}$ ($t=1.373$),
 $10^{9.12}$ ($t=1.374$), and
 $10^{10.06}$ ($t=1.374$).

{\bf Fig.10.} --- ({\em a}) \& ({\em b}):
 snapshots of Model D ($t=1.374$ \& $\rho_c=10^{10}\rho_s$),
 which is similar to model C but the critical density
 above which the equation of state is hard is assumed $10^6 \rho_s$.
Comparing level 9 ({\em a}) and level 11 ({\em b}),
 it is shown that two shock fronts exist: the outer one is
 seen in ({\em a}) in a distance of $r\simeq 10^{-3}$ from the center
 of the cloud core.
Another one is seen in ({\em b}) at $z\simeq 1.5\times 10^{-4}$ near
 the $z$-axis and $r\simeq 3.5\times 10^{-4}$ and $z\sim 0$.
({\em c}) \& ({\em d}): cross-cut views along the $z$- and $r$-axis,
 respectively.
Snapshots correspond to $\rho_c=10^6$ ($t=1.372$),
 $10^{7.1}$ ($t=1.373$), $10^{8.21}$ ($t=1.374$),
 $10^9$, ($t=1.374$) \&
 $10^{10.05}$ ($t=1.374$)
\clearpage

\end{document}